# Four-wave interaction in gas and vacuum. Definition of a third order nonlinear effective susceptibility in vacuum : $\chi^{(3)}_{vacuum}$

F.Moulin[a*] , D.Bernard[b]

[a]Département de Physique, Ecole Normale Supérieure de Cachan, 94235 Cachan, France 61 av du prés Wilson, 94235 Cachan, France
[b]LPNHE, Ecole Polytechnique, IN2P3&CNRS, 91128 Palaiseau, France

[*] **e-mail:** moulin@physique.ens-cachan.fr

**Abstract :** Semiclassical methods are used to study the nonlinear interaction of light in vacuum in the context of four wave mixing. This study is motivated by a desire to investigate the possibility of using recently developed powerful ultrashort (femtosecond) laser pulses to demonstrate the existence of nonlinear effects in vacuum, predicted by quantum electrodynamics (QED). An approach, similar to classical nonlinear optics in a medium, is developed in this article. A third order nonlinear effective susceptibility of vacuum is then introduced .

## 1. Introduction

It has been known for some time now that quantum electrodynamics (QED) predicts the existence of a nonlinear interaction between electromagnetic fields in vacuum [1-7].

The effects caused by the vacuum polarization are various. Electric or magnetic anisotropy of vacuum, are the subject of interesting theoretical and experimental research. For example the PVLAS [8,9] is a experiment designed to measure the vacuum magnetic birefringence. It is based on a very sensitive ellipsometer and on the use of a strong magnetic field and weak laser beams.

In our laboratory, progress in powerful ultrashort (femtosecond) laser pulses [10] is an opportunity for studying QED phenomena in intense laser beams electromagnetic fields. The four wave nonlinear optical mixing process (FWM) in vacuum seems to be also a interesting way to study vacuum nonlinearities [11-14]. In the present work, we analyze the achievability of FWM in vacuum, and then we compare these theoretical results with FWM in a medium. A third order nonlinear effective susceptibility of vacuum, $\chi^{(3)}_{vacuum}$, is then introduced to satisfy the same differential equation as the one obtained in a gas characterized by an effective $\chi^{(3)}_{gas}$. The introduction of $\chi^{(3)}_{vacuum}$ is a new and easy way to compute the number of generated photons in vacuum. In the last part of this article, a numerical estimation of the theoretical power required to stimulate one photon by FWM with an ultrashort laser pulse is performed. The power of existing laser pulses [10] is only a few orders of magnitude from this theoretical estimation. Nevertheless, an experimental program is underway at Ecole Polytechnique to search for possible non-QED new physics in photon-photon elastic scattering at low photon energies (electron volts). A two beam experiment was performed in 1995, where an IR and a green laser beam were brought to a head-on collision in vacuum, and possible scattered photons were searched at angle [15]. A FWM experiment was also performed recently in 1997 [16] with three focused beams crossing each other in a three-dimensional geometry. A new two-dimensional FWM experiment is proposed with two beams at the same wavelength ( $\lambda_1=\lambda_2=800$ nm) and a third beam generated by a parametric oscillator at $\lambda_3=1300$ nm. Stimulated photons at $\lambda_4=577$ nm were expected in gas or perhaps in vacuum.

## 2. Fundamental equations

We start with the Lagrange-function density $\Lambda$, including the Euler-Heisenberg radiation correction term, $\delta\Lambda$, induced by vacuum polarization [1] :

$$\Lambda = \Lambda_0 + \delta\Lambda = \Lambda_0 + a\left[(\vec{E}^2 - c^2\vec{B}^2)^2 + 7c^2(\vec{E}.\vec{B})^2\right]$$

with $\quad a = \dfrac{2\alpha^2 \hbar^3 \varepsilon_0^2}{45 m^4 c^5} = \dfrac{\hbar e^4}{360\pi^2 m^4 c^7}$

$\Lambda_0 = \dfrac{\varepsilon_0}{2}(\vec{E}^2 - c^2\vec{B}^2)$ is the standard Lagrangian-function density of electromagnetism, $\alpha$ the fine structure constant, e and m the electron charge and mass, $\hbar$ the Planck's constant and c the velocity of light.

Such a lagrangian is used for relatively low-frequency electromagnetic field, as optical laser beams $\omega \ll \omega_c = \dfrac{mc^2}{2\hbar}$.

By varying the action $S = \iint \Lambda \, dt \, d\tau$ we obtain Maxwell's equations [17] :

$$\vec{\nabla}.\vec{B} = 0 \qquad\qquad \vec{\nabla} \wedge \vec{E} = -\dfrac{\partial \vec{B}}{\partial t}$$

$$\vec{\nabla}.\vec{E} = -\dfrac{\vec{\nabla}.\vec{P}}{\varepsilon_0} \qquad\qquad \vec{\nabla} \wedge \vec{B} = \mu_0 \left( \vec{\nabla} \wedge \vec{M} + \dfrac{\partial \vec{P}}{\partial t} + \varepsilon_0 \dfrac{\partial \vec{E}}{\partial t} \right)$$

with $\vec{P}$ and $\vec{M}$, the nonlinear vacuum electric and magnetic polarization vectors, given by [17]: $\vec{P} = \dfrac{\partial \delta\Lambda}{\partial \vec{E}}$ and $\vec{M} = \dfrac{\partial \delta\Lambda}{\partial \vec{B}}$ with the notation : $\dfrac{\partial}{\partial \vec{E}} = \dfrac{\partial}{\partial E_x}\vec{i} + \dfrac{\partial}{\partial E_y}\vec{j} + \dfrac{\partial}{\partial E_z}\vec{k}$

Finally we obtain [11] :

$$\begin{cases} \vec{P} = 2a[2(\vec{E}^2 - c^2\vec{B}^2)\vec{E} + 7c^2(\vec{E}.\vec{B})\vec{B}] \\ \vec{M} = 2a[7c^2(\vec{E}.\vec{B})\vec{E} - 2c^2(\vec{E}^2 - c^2\vec{B}^2)\vec{B}] \end{cases} \qquad (1)$$

With Maxwell's equations, we obtain the propagation equation of the electric field:

$$\Delta\vec{E} - \dfrac{1}{c^2}\dfrac{\partial^2 \vec{E}}{\partial t^2} = \mu_0 \left[ \dfrac{\partial}{\partial t}\vec{\nabla} \wedge \vec{M} + \dfrac{\partial^2 \vec{P}}{\partial t^2} - c^2\vec{\nabla}(\vec{\nabla}.\vec{P}) \right] \qquad (2)$$

Equation (2) is also valid in a nonlinear medium with different $\vec{P}$ and $\vec{M}$ expressions including linear and nonlinear terms.

## 3. Calculation of four-wave mixing

### 3-1 Definitions and initial equation

We consider three electromagnetic waves $(\vec{E}_1, \vec{B}_1)$, $(\vec{E}_2, \vec{B}_2)$, $(\vec{E}_3, \vec{B}_3)$ crossing each other in a nonlinear media. A wave $(\vec{E}_4, \vec{B}_4)$ is generated by a four wave difference frequency mixing with the wavelength and direction given by the energy-momentum conservation conditions :

$$\begin{cases} \vec{k}_4 = \vec{k}_1 + \vec{k}_2 - \vec{k}_3 \\ \omega_4 = \omega_1 + \omega_2 - \omega_3 \end{cases} \qquad (3)$$

The deviation from the exact synchronization condition is negligible if $\Delta k.L \ll 1$ with L the interaction length and $\Delta \vec{k} = \vec{k}_4 - (\vec{k}_1 + \vec{k}_2 - \vec{k}_3)$. For crossed light beams in vacuum or in low gas pressure we consider that the phase matching condition is satisfied : $\Delta \vec{k} = \vec{0}$.

In the general case of linearly polarized beams, we introduce the complex electric and magnetic fields :

$$\vec{E}_j = \frac{1}{2} \{ \underline{E}_{0j}(\vec{r},t) \vec{u}_j e^{i(\omega_j t - \vec{k}_j \cdot \vec{r})} + c.c \} \quad j=1,2,3,4$$

$$\vec{B}_j = \frac{1}{2} \{ \underline{B}_{0j}(\vec{r},t) \vec{v}_j e^{i(\omega_j t - \vec{k}_j \cdot \vec{r})} + c.c \}$$

with $\vec{u}_j$ and $\vec{v}_j$ defined as unit vectors.

In the relations (1), the electric and magnetic fields, $\vec{E}$ and $\vec{B}$, are the superposition of the four fields $\vec{E} = \vec{E}_1 + \vec{E}_2 + \vec{E}_3 + \vec{E}_4$ and $\vec{B} = \vec{B}_1 + \vec{B}_2 + \vec{B}_3 + \vec{B}_4$. In vacuum or in low pressure gas we suppose that the relations $\omega_j = k_j c$ and $c\underline{B}_{0j} = \underline{E}_{0j}$ hold.

We define the z axis as the direction $\vec{k}_4$ of the generated field $(\vec{E}_4, \vec{B}_4)$. The third order components of $\vec{P}$ and $\vec{M}$ at $(\omega_4, \vec{k}_4)$, noted $\vec{P}_4$ and $\vec{M}_4$, are :

$$\vec{P}_4(\vec{r},t) = \frac{1}{2} \{ \underline{\vec{P}}_{04}(\vec{r},t) e^{i(\omega_4 t - k_4 z)} + c.c \} \qquad (4)$$

$$\vec{M}_4(\vec{r},t) = \frac{1}{2} \{ \underline{\vec{M}}_{04}(\vec{r},t) e^{i(\omega_4 t - k_4 z)} + c.c \}$$

We regard the amplitude of $E_{04}$, $B_{04}$, $P_{04}$, $M_{04}$ as slowly varying functions of the coordinate z and t (on the scale of the radiation wavelength and period). The fourth field is assumed to propagate concentrically along the z axis. The differential equation (2), for the amplitude components $\underline{E}_{04}$, assumes the form:

$$\left[\frac{c}{2i\omega_4}\Delta_\perp \underline{E}_{04} - \frac{\partial \underline{E}_{04}}{\partial z} - \frac{1}{c}\frac{\partial \underline{E}_{04}}{\partial t}\right]\vec{u}_4 = \frac{i\mu_0\omega_4}{2}[(c\underline{P}_{04x} + \underline{M}_{04y})\vec{u}_x + (c\underline{P}_{04y} - \underline{M}_{04x})\vec{u}_y] \quad (5)$$

with $\Delta_\perp = \frac{\partial^2}{\partial x^2} + \frac{\partial^2}{\partial y^2}$ the transverse Laplacian.

This equation describes the growth of $\underline{E}_{04}$, both in a medium and in vacuum where the phase matching condition is satisfied.

### 3-2 Influence of field derivative terms in Lagrangian

A modified version of the Heisenberg-Euler theory in which the Lagrangian incorporates terms with field derivates, can be used to take into account vacuum dispersion. In our FWM experiment we show that the influence of these terms is negligible.

In the lowest approximation in dispersion and nonlinearity we have [12-14] :

$$\begin{cases} \vec{P} = 2a[2(\vec{E}^2 - c^2\vec{B}^2)\vec{E} + 7c^2(\vec{E}.\vec{B})\vec{B}] - 6g\left(\frac{1}{c^2}\frac{\partial^2 \vec{E}}{\partial t^2} - \Delta\vec{E}\right) \\ \vec{M} = 2a[7c^2(\vec{E}.\vec{B})\vec{E} - 2c^2(\vec{E}^2 - c^2\vec{B}^2)\vec{B}] + 6gc^2\left(\frac{1}{c^2}\frac{\partial^2 \vec{B}}{\partial t^2} - \Delta\vec{B}\right) \end{cases}$$

with $g = \frac{e^2\hbar}{360\pi m^2 c^3}$.

We study the influence of a intense electromagnetic field, $(\vec{E}_0, \vec{B}_0)$, on the propagation of a pulse laser beam $(\vec{E}_1, \vec{B}_1)$. The interaction of two waves, rather than three, satisfied the synchronization conditions if : $\vec{k}_2 = \vec{k}_3$, $\omega_2 = \omega_3$, $\vec{E}_2 = \vec{E}_3 = \vec{E}_0$, $\vec{B}_2 = \vec{B}_3 = \vec{B}_0$

For simplicity we consider the case of two oppositely directed waves with the same linear polarization states. A small deviation of the refractive index from unity, $\delta n$, is calculated at a fixed polarization in the absence of dispersion (g=0). Using Maxwell equations we obtain the dispersion relation :

$$\frac{k_1}{\omega_1}=\frac{1}{c}[1+\delta n+12\mu_0 g\omega_1^2(\delta n)^2] \quad \text{with} \quad \delta n=\frac{2a}{\varepsilon_0}|E_0+cB_0|^2=\frac{8a}{\varepsilon_0}|E_0|^2$$

The spread of pulse is characterized by the quadratic dispersion parameter $D=\frac{\partial^2 k_1}{\partial \omega_1^2}$ [12-14], and by the characteristic dispersion length :

$$L_{disp}=\frac{\tau^2}{D}=\frac{c\tau^2}{72\mu_0 g\omega_1(\delta n)^2}=10^{39}\tau^2\frac{\omega_c}{\omega_1}(\frac{E_c}{E_0})^4$$

with $\tau$ the pulse duration and $E_c=\frac{m^2c^3}{e\hbar}$ the critical field at which electron-positron pairs are efficiently produced.

For $\frac{\omega_c}{\omega_1}\approx 10^5$, $\frac{E_c}{E_0}\approx 10^4$ and $\tau\approx 30 fs$ we obtain $L_{disp}\approx 10^{33}m$, which is much larger than typical interaction length in a crossed beams geometry.

We see that vacuum dispersion does not occur in FWM experiment. Self-action of radiation or other nonlinear optics effects in vacuum are also negligible in this case.

### 3-3 Third order nonlinear effective susceptibility in a gas : $\chi_{gas}^{(3)}$

Because of the inversion symmetry in gas only odd order interactions are allowed through electric dipole coupling. We study FWM in a low pressure gas where the phase matching condition is satisfied. In a non magnetic case, $\underline{M}_{04x}$ and $\underline{M}_{04y}$ are equal to zero in equation (5). The electric polarizations, $\underline{P}_{04x}$ and $\underline{P}_{04y}$, are related to a third order nonlinear susceptibility tensor $\underline{\underline{\chi}}^{(3)}$ by the relations :

$$\underline{P}_{04x}(\omega_4)=\varepsilon_0\sum_{k,l,m}\chi_{xklm}^{(3)}(\omega_4)\underline{E}_{01k}(\omega_1)\underline{E}_{02l}(\omega_2)\underline{E}_{03m}^*(\omega_3) \tag{6}$$

$$\underline{P}_{04y}(\omega_4)=\varepsilon_0\sum_{k,l,m}\chi_{yklm}^{(3)}(\omega_4)\underline{E}_{01k}(\omega_1)\underline{E}_{02l}(\omega_2)\underline{E}_{03m}^*(\omega_3)$$

$\chi_{jklm}^{(3)}$ is a tensor element for the amplitude component j=x,y or z of the polarization vector. The indices k,l,m represent the three directions of polarization (x,y or z) for each of the three incident fields.

It is common to define an effective susceptibility in gas, $\chi^{(3)}_{gas}$, as a linear combination of $\chi^{(3)}_{jklm}$, which depends only on the laser beam polarization :

$$\vec{P}_{04}(\omega_4) = \underline{P}_{04x}\vec{u}_x + \underline{P}_{04y}\vec{u}_y = \varepsilon_0 \chi^{(3)}_{gas}(\omega_4)\underline{E}_{01}(\omega_1)\underline{E}_{02}(\omega_2)\underline{E}^*_{03}(\omega_3)\vec{u}_4$$

With this notation equation (5) reduces to:

$$\frac{c}{2i\omega_4}\Delta_\perp \underline{E}_{04} - \frac{\partial \underline{E}_{04}}{\partial z} - \frac{1}{c}\frac{\partial \underline{E}_{04}}{\partial t} = \frac{i\omega_4}{2c}\chi^{(3)}_{gas}\underline{E}_{01}\underline{E}_{02}\underline{E}^*_{03} \qquad (7)$$

## 3-4 Third order nonlinear effective susceptibility in vacuum : $\chi^{(3)}_{vacuum}$

We want to introduce an effective susceptibility in vacuum, $\chi^{(3)}_{vacuum}$, which satisfies the same equation as (7), obtained in a low pressure gas. Relations (1) and (4) give :

$$\vec{P}_{04} = 2a\underline{E}_{01}\underline{E}_{02}\underline{E}^*_{03}\vec{K}_P$$

$$\vec{M}_{04} = 2ac\underline{E}_{01}\underline{E}_{02}\underline{E}^*_{03}\vec{K}_M$$

with the geometric factors $\vec{K}_P$ and $\vec{K}_M$ given by :

$$\begin{aligned}\vec{K}_P &= \vec{u}_1(\vec{u}_2.\vec{u}_3 - \vec{v}_2.\vec{v}_3) + \vec{u}_2(\vec{u}_1.\vec{u}_3 - \vec{v}_1.\vec{v}_3) + \vec{u}_3(\vec{u}_1.\vec{u}_2 - \vec{v}_1.\vec{v}_2) + \\ &\quad 7/4[\vec{v}_1(\vec{u}_2.\vec{v}_3 + \vec{u}_3.\vec{v}_2) + \vec{v}_2(\vec{u}_1.\vec{v}_3 + \vec{u}_3.\vec{v}_1) + \vec{v}_3(\vec{u}_1.\vec{v}_2 + \vec{u}_2.\vec{v}_1)] \\ \vec{K}_M &= \vec{v}_1(\vec{v}_2.\vec{v}_3 - \vec{u}_2.\vec{u}_3) + \vec{v}_2(\vec{v}_1.\vec{v}_3 - \vec{u}_1.\vec{u}_3) + \vec{v}_3(\vec{v}_1.\vec{v}_2 - \vec{u}_1.\vec{u}_2) + \\ &\quad 7/4[\vec{u}_1(\vec{u}_2.\vec{v}_3 + \vec{u}_3.\vec{v}_2) + \vec{u}_2(\vec{u}_1.\vec{v}_3 + \vec{u}_3.\vec{v}_1) + \vec{u}_3(\vec{u}_1.\vec{v}_2 + \vec{u}_2.\vec{v}_1)]\end{aligned} \qquad (8)$$

Assuming the same differential equation (7) for vacuum and gas, we obtain :

$$\begin{aligned}\chi^{(3)}_{vacuum} &= \frac{2a}{\varepsilon_0}\left[(K_{Px} + K_{My})^2 + (K_{Py} - K_{Mx})^2\right]^{1/2} \\ &= \frac{2aK}{\varepsilon_0} = \frac{2\hbar e^4 K}{360\pi^2 m_e^4 c^7 \varepsilon_0} = 2.92 \, 10^{-41} K \, (m^2/V^2)\end{aligned} \qquad (9)$$

with $K = \|\vec{K}\|$ and $\vec{K} = (K_{Px} + K_{My})\vec{u}_x + (K_{Py} - K_{Mx})\vec{u}_y$

The wave vector polarization of the generated field is $\vec{u}_4 = \vec{K}/K$

K is maximum for degenerate four-wave mixing (phase conjugation) with a value of 14 ($\chi^{(3)}_{vacuum} = 4.10^{-40} \, m^2/V^2$).

In contrast with the usual problems of nonlinear optics, the vacuum is characterized by both the electric and the magnetic nonlinear polarizations simultaneously. In opposition to the gas case, $\chi^{(3)}_{vacuum}$ depends both on the laser polarization and also on the laser beam geometry. It is not possible to use here a nonlinear susceptibility tensor. Therefore, for each experimental configuration an effective scalar susceptibility will be easily calculated.

**3-5  Critical gas pressure**

A critical gas pressure, $p_{cr}$, can be defined when the number of generated photons is equal in vacuum and in gas. It is, of course, realized when $\chi^{(3)}_{vacuum} = \chi^{(3)}_{gas}$. With the relation :

$$\chi^{(3)}_{gas} = \frac{p\gamma^{(3)}_{gas}}{kT\varepsilon_0} \quad \text{we obtain} \quad p_{cr} = \frac{\varepsilon_0 kT \chi^{(3)}_{vacuum}}{\gamma^{(3)}_{gas}} .$$

k is the Boltzman constant, $\gamma^{(3)}_{gas}$ the third order nonlinear hyperpolarisability of a gas molecule, and T the temperature in Kelvin .

For example, if we consider nitrogen (N$_2$) gas with a typical $\gamma^{(3)}_{N_2} \approx 10^{-63} Cm^4/V$ [18], for T=300 Kelvin and $\chi^{(3)}_{vacuum} = 4.10^{-40} m^2/V^2$, we obtain $p_{cr} \approx 10^{-10} mbar$.

**3-6  $\chi^{(3)}_{vacuum}$ in 2D geometry**

**3-6-1  definitions**

Relations (3) are used to obtain the geometric configuration of the four beams. $\lambda_1$, $\lambda_2$, $\lambda_3$ and $\alpha_2$ are given parameters, and $\lambda_4$, $\alpha_3$, $\alpha_4$ are computed using relations (3), with the angles defined in figure 1:

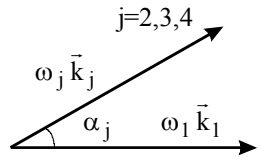

**Fig 1 :** definition of the angle of the four beams. Beam 1 defines the reference axis.

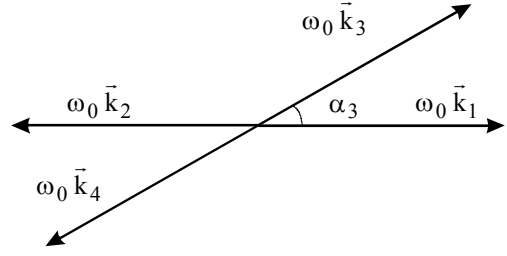

**Fig 2 :** Geometric configuration for optical phase conjugation.

### 3-6-2  Optical phase conjugation

This case is obtained for a degenerate four-wave mixing $\omega_1 = \omega_2 = \omega_3 = \omega_4 = \omega_0$. The energy-momentum conservation condition (3), is verified in this case only for a configuration with at least two head-on laser beams as shown in figure 2 .

The fourth beam is then generated at the opposite direction of the third laser beam at $\alpha_4 = \alpha_3 + \pi$ , with the same wavelength.

The K constant depends on laser polarization and angle $\alpha_3$. We study two cases. The first one $K_{//}$, is obtained when the polarization vectors $\vec{u}_1$ , $\vec{u}_2$, $\vec{u}_3$ are in the same direction, and the second case $K_\perp$, when the polarization vectors $\vec{u}_1$, $\vec{u}_3$ are in the same direction and $\vec{u}_2$ perpendicular to $\vec{u}_1$ ($\vec{u}_2 = \vec{v}_1$).

With relations (8) and (9) we obtain : $K_{//} = 2(3 + \cos^2 \alpha_3)$ and $\vec{u}_{4//} = \vec{u}_3$. $K_{//} = 8$ is the maximum value for $\alpha_3 = 0$.

$K_\perp = \dfrac{1}{4}((22 + 3\cos\alpha_3)\cos\alpha_3 + 31)$ and $\vec{u}_{4\perp} = \vec{v}_3$   $K_\perp = 14$ is the maximum value for $\alpha_3 = 0$.

### 3-6-3  Non degenerate four-wave mixing

Optical phase conjugation uses four beams at the same wavelength. This experimental set-up is nearly impossible to realize because of the difficulties of detecting some generated photons in an intense noise background at the same wavelength. A two-dimensional set-up is proposed to detect generated photons with a different wavelength and different angle. We study the interesting case of two beams with the same wavelength $\lambda_1 = \lambda_2$ and a third beam at $\lambda_3$.

For example with $\lambda_1=\lambda_2=\lambda_0=800$nm and $\lambda_3=1300$nm we obtain $\lambda_4= 577$nm. In figure 3 $\alpha_3$, $\alpha_4$ and K (for $\vec{u}_1 = \vec{u}_3$ and $\vec{u}_2 = \vec{v}_1$) are obtained versus $\alpha_2$. Solutions are available for angles $-\alpha_{2c} < \alpha_2 < \alpha_{2c}$ with $\cos\alpha_{2c} = 2(1-\frac{\lambda_0}{\lambda_3})^2 - 1$. In figure 3 $\alpha_{2c} = 137.8°$:

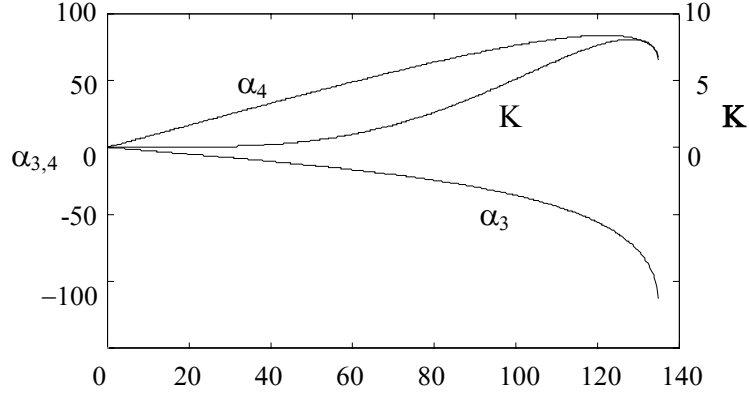

**Fig 3 :** Angle values for beam 4 (stimulated beam) and laser beam 3, versus $\alpha_2$ in degrees, with $\lambda_1=\lambda_2=800$nm and $\lambda_3=1300$nm. K is the geometric factor.

### 3-7  Number of generated photons

We now calculate the number of photons generated by FWM in a gas or vacuum, by solving equation (7) for different laser profiles and for the plane wave approximation.

### 3-7-1  Plane wave approximation

With the plane wave approximation, equation (7) reads :

$$\frac{d\underline{E}_{04}}{dz} = -\frac{i\omega_4}{2c} \chi^{(3)} \underline{E}_{01} \underline{E}_{02} \underline{E}_{03}^* \qquad (10)$$

We define an interaction length L and we assume that the electric field $E_{04}$ is equal to zero at $z=-L/2$, and becomes maximum at $z=+L/2$. After integration over the interaction length, L, we obtain the maximum electric field at $z=L/2$ :

$$\underline{E}_{04}(z = L/2) = -\frac{i\omega_4 L}{2c} \chi^{(3)} \underline{E}_{01} \underline{E}_{02} \underline{E}_{03}^* \qquad (11)$$

We obtain the number of photons $N_4$ generated by four wave mixing by integration :

$$N_4 = \frac{\pi\varepsilon_0 c}{\hbar\omega_4} \int_0^{+\infty} \int_{-\infty}^{+\infty} \underline{E}_{04} \underline{E}_{04}^*(z = L/2) \rho \, d\rho \, dt \qquad (12)$$

### 3-7-2 Parallel gaussian beams

First we assume three Gaussian collinear pump pulses :

$$E_{0j}(\rho,t)=E_{0j}(0)\exp(-\frac{\rho^2}{w_{0j}^2}-\frac{t^2}{\tau^2}) \quad \text{for } j=1,2,3$$

with $w_{0j}$ the beam waist at 1/e field radius, and $\tau$ the pulse duration. $\rho=\sqrt{x^2+y^2}$

Relation (12) gives :

$$N_4=(\frac{2}{\pi\sqrt{3}})^3 \frac{3\omega_4(\chi^{(3)})^2 L^2 \varepsilon_1\varepsilon_2\varepsilon_3}{\hbar\varepsilon_0^2 c^4 \tau^2(w_{01}^2 w_{02}^2 + w_{01}^2 w_{03}^2 + w_{02}^2 w_{03}^2)} \tag{13}$$

with $\varepsilon_j=\frac{c\tau\varepsilon_0 E_{0j}^2(0)w_{0j}^2}{2}(\frac{\pi}{2})^{3/2}$ the incident laser beam energy. Relation (13) is only valid when $c\tau > L$.

With ultrashort pulses, $c\tau < L$, we use $L=c\tau$ in relation (13). A numeric form for (13) in the vacuum case is :

$$(N_4)_{vacuum}=3.510^{-3} K^2 \frac{L^2(\mu m)\varepsilon_1(J)\varepsilon_2(J)\varepsilon_3(J)}{\lambda_4(\mu m)\tau^2(fs)(w_{01}^2(\mu m)w_{02}^2 + w_{01}^2 w_{03}^2 + w_{02}^2 w_{03}^2)} \tag{14}$$

### 3-7-3 Gaussian collinear focused beams

We consider now three Gaussian collinear pump beams focused to a common point at z=0 :

$$\underline{E}_{0j}(\rho,z,t)=\frac{E_{0j}(0)}{(1-2iz/b)}\exp(-\frac{\rho^2}{w_{0j}^2(1-2iz/b)}-\frac{t^2}{\tau^2})=E_{0j}(0)\underline{E}'_{0j}(\rho,z)\exp(-\frac{t^2}{\tau^2})$$

with j=1,2,3

b is the confocal parameter (for diffraction limited beams we have $b=k_j w_{0j}^2$).

The resolution of equation (10) gives the same relation (13) with L= $L_{eq}$, where $L_{eq}$ is an equivalent interaction length defined by :

$$L_{eq}^2 = 4(1/w_{01}^2 + 1/w_{02}^2 + 1/w_{03}^2)\int_0^{+\infty} \rho \underline{JJ}^* d\rho$$

with $\underline{J}(\rho) = \int_{-\infty}^{+\infty} \underline{E}'_{01}\underline{E}'_{02}\underline{E}'^{*}_{03}(\rho;z)dz$ \quad (15)

In the collinear approximation, $L_{eq}$ is often close to b.

A review of effects of focusing on third-order nonlinear processes in isotropic media is done by Bjorklund [19]

### 3-7-4 Non collinear Gaussian focused beams

The laser beams propagate in different directions with different wavelengths. Therefore, the collinear approximation is not valid. A numerical program is made to obtain an equivalent interaction length, $L_{eq}$, by using Gaussian beams with the following expression :

$$\underline{E}_{0j}(x_j, y_j, z_j, t) = \frac{E_{0j}(0)}{(1 - 2iz_j/b)} \exp\left(-\frac{x_j^2 + y_j^2}{w_{0j}^2(1 - 2iz_j/b)} - \frac{t^2}{\tau^2}\right)$$ where $x_j$, $y_j$, $z_j$ are functions of

x, y, z, the coordinates of the generated beam. In the following section the equivalent length is computed with non collinear beams. In the non collinear approximation, $L_{eq}$ is often close to $w_0$.

### 3-7-5 Numerical applications

In the case of $\lambda_1=\lambda_2=800$nm, $\lambda_3=1300$nm, $\lambda_4=577$nm an experimental 2D configuration is obtained, as in figure (3), for $\alpha_2=110°$. Then we compute $\alpha_3=-44°$, $\alpha_4=81°$ and K=6.5.
For ultrashort laser pulses ($\tau=30$fs) and in the diffraction limit with focal spots $w_{01}=w_{02}=w_{03}=5\mu m$, we compute $L_{eq}=4.4\mu m$. We then need $\varepsilon_1 \varepsilon_2 \varepsilon_3 = 5.10^5$ $J^3$ or $\varepsilon_1 = \varepsilon_2 = \varepsilon_3 = 80$ J to obtain one photon per laser shot. Using a 10Hz repetition rate laser and with beam energies of $\varepsilon_1=\varepsilon_2=12$J, $\varepsilon_3=100$ mJ, we obtain one photon per hour.

### 3-7-6 FWM experiment

A three beam experiment with ultrashort laser pulses ($\tau=40$fs) was performed recently [16]. Two laser beams at 805nm produced by a Ti:Saphire laser, and a third beam at 1300nm generated by a parametric oscillator, were brought to collision in vacuum. The three beams were focused by a single optic made of a pair of spherical mirrors (Bowen) in a 3-dimensional configuration. Stimulated photons at $\lambda_4 \approx 561$nm were observed in a gas jet but not in vacuum. The parameters of the experiment were : $w_{01}=w_{02}=4\mu m$, $w_{03}=6\mu m$, $\varepsilon_1=15$mJ, $\varepsilon_2=5.5$mJ $\varepsilon_3=20\mu J$. The numerical calculation gives $K \approx 0.56$ and $L_{eq} \approx 5$ $\mu m$. Taking into account loss factors (photomultiplier efficiency, spectrometer transmission ...), we obtain finally a theoretical prediction of $(N_4)_{vacuum} \approx 6.2 \ 10^{-20}$ photons per shot while the observed limit is 6 10$^{-3}$ photons per shot. The experimental result is 17 order of magnitude under QED prediction. Several orders of magnitude could be gained by an improvement in the tuning of the laser, the OPA, and by further work on the background noise.

## 4. Conclusion

We have compared the calculation of four wave mixing in vacuum and in a gas. The identical form of the equation giving the growth rate of the fourth beam allowed us to define a third order nonlinear susceptibility in vacuum, $\chi^{(3)}_{vacuum}$. This susceptibility is computed for a geometric laser configuration in two or three dimensions. Some experimental arrangements are proposed to eventually observe four wave mixing in vacuum with powerful laser beams. Numerical estimations are made with ultrashort laser pulses. For instance, the purpose of the experimental program underway at Ecole Polytechnique, is to search for possible non-QED new physics at low photon energies, but, with the evolution of laser technology, the QED effect predicted in low energy will probably be observed in a few years.

**Acknowlegments**

We gratefully acknowledge the help of the staff of the LOA and LULI for technical assistance during the experiment. We thank also F.Amiranoff and A.Braun for the correction of the manuscript.

**Réferences**

[1]   H.Euler, Ann. der Phys. (Leipzig) 26 (1936) 398

[2]   W.Heisenberg and H.Euler, Z.Physik 98 (1936) 714

[3]   R.Karplus and M.Neuman, Phys. Rev 80 (1950) 380

[4]   R.Karplus and M.Neuman, Phys.Rev 83 (1951) 776

[5]   N.Kroll, Phys. Rev. 127 (1962) 1207

[6]   J.Mc Kenna and P.M Platzman, Phys. Rev 129 (1963) 2354

[7]   A.A Varfolomeev, JETP 23 (1966) 681

[8]   D.Bakalov et al., "Experimental method to detect the magnetic birefringence of vaccum" , Quantum Semiclass. Opt., 1998, vol. 10, pp 239-250

[9]   D.Bakalov et al., Nuclear Physics B (Proc. Suppl. ) 35(1994) 180-182 North-holland

[10]  A.Antonetti, and al., Appl. Phys. B 65 (1997) 197

[11]  G. Grynberg and J.Y Courtois, C.R. Acad. Sci. Paris t.311, Série II (1990) 1149

[12]  N.N Rozanov, JETP vol. 76, No 6 (1993) 991.

[13]  N.N Rozanov, JETP vol. 86, No2 (1998).

[14]  N.N Rozanov, S.V. Fedorov Optics and Spectr. Vol. 84, No1 (1998)

[15]  F.Moulin, D.Bernard, F.Amiranoff, Z.Phys. C 72 (1996) 607

[16]  D.Bernard, F.Moulin et al, "Search for stimulated Photon-Photon scattering in vaccum" . European Journal D, 10 (2000) 141

[17]  V.B Berestetskii, E.M Lifshitz, L.P Pitaevskii, *Quantum Elect. (Moscow, 1989)*

[18]  H.J.Lehmeier, W.Leupacher, A.Penzkofer, Optics communications 56 (1985) 67

[19]  G.C.Bjorklund, IEEE Journal of Quantum Elect. Vol. QE-11 No.6 (June 1975)